\documentclass[fleqn,usenatbib]{mnras}

\usepackage{newtxtext,newtxmath}

\usepackage{adjustbox}
\usepackage{multirow} 
\usepackage{array,makecell}

\usepackage[T1]{fontenc}
\usepackage{ae,aecompl}

\usepackage[nameinlink]{cleveref}
\usepackage[utf8]{inputenc}
\usepackage{graphicx,float}	
\usepackage{adjustbox} 
\usepackage{amsmath}	
\usepackage{stackengine}
    \newcommand\xrowht[2][0]{\addstackgap[0.5\dimexpr#2\relax]{\vphantom{#1}}}
\restylefloat{table}

\usepackage{subcaption}
\usepackage{lipsum}
\usepackage{arydshln}
\usepackage{ragged2e}

\usepackage{ulem}

\title[Mass Derivation of planets K2-21b and K2-21c from Transit Timing Variations]{Mass Derivation of planets K2-21b and K2-21c from Transit Timing Variations}

\author[El Moutamid et al.]{Maryame El Moutamid$^{1}$\thanks{E-mail: maryame@astro.cornell.edu},
Kevin, B. Stevenson$^{2},$
Billy Quarles$^{3,4},$
Nikole, K. Lewis$^{1},$
\newauthor Erik Petigura$^{5},$
Daniel Fabrycky$^{6},$
Jacob L.\ Bean$^{6},$
Diana Dragomir$^{7},$
Kristin S. Sotzen$^{2}$ 
\newauthor and Michael, W. Werner$^{8}$\\ \\
$^{1}$Cornell Center of Astrophysics and Planetary Sciences, Department of Astronomy and Carl Sagan Institute, Ithaca, NY, USA\\
$^2$Johns Hopkins Applied Physics Laboratory, Laurel, MD, USA \\
$^3$Center for Relativistic Astrophysics, School of Physics, Georgia Institute of Technology, Atlanta, GA 30332, USA \\
$^4$Department of Physics, Astronomy, Geosciences and Engineering Technology, Valdosta State University, Valdosta, GA 31698, USA \\
$^{5}$Department of Physics and Astronomy, University of California Los Angeles, Los Angeles, CA, USA \\
$^{6}$Department of Astronomy and Astrophysics, University of Chicago, Chicago, IL, USA \\
$^{7}$Department of Physics and Astronomy, University of New Mexico, Albuquerque, NM, USA \\
$^{8}$Jet Propulsion Laboratory, California Institute of Technology, Pasadena, CA, USA
}

\date{Accepted XXX. Received YYY; in original form ZZZ}

\pubyear{2022}

\begin{document}

\label{firstpage}
\pagerange{\pageref{firstpage}--\pageref{lastpage}}
\maketitle

\begin{abstract} 

While various indirect methods are used to detect exoplanets, one of the most effective and accurate methods is the transit method, which measures the brightness of a given star for periodic dips when an exoplanet is passing in front of the parent star.
For systems with multiple transiting planets, the gravitational perturbations between planets affect their transit times. 
The difference in transit times allows a measurement of the planet masses and orbital eccentricities. These parameters help speculating on the formation, evolution and stability of the system. Using Transit Timing Variations (TTVs), we measure the masses and eccentricities of two planets orbiting K2-21, a relatively bright K7 dwarf star. These two planets exhibit measurable TTVs, have orbital periods of about $9.32$ days and $15.50$ days, respectively, and a period ratio of about $1.66$, which is relatively near to the 5:3 mean motion resonance. We report that the inner and outer planets in the K2-21 system have properties consistent with the presence of a hydrogen and helium dominated atmospheres, as we estimate their masses to be $1.59^{+0.52}_{-0.44}\ M_\oplus$ and $ 3.88^{+1.22}_{-1.07}\ M_\oplus$ and densities of $ 0.22^{+0.05}_{-0.04}\ \rho_\oplus$ and $0.34^{+0.08}_{-0.06}\ \rho_\oplus$, respectively ($M_\oplus$ and $\rho_\oplus$ are the mass and density of Earth, respectively).  
Our results show that the inner planet is less dense than the outer planet; one more counter-intuitive exoplanetary system such as Kepler-105, LTT 1445, TOI-175 and Kepler-279 systems. 


\end{abstract}




\section{Introduction} \label{sec_introduction}

The Kepler mission revealed that the most numerous type of planet in our galaxy has a size in between that of Earth and Neptune \citep{Howard2012,Petigura2013,Dressing2015a}. Without solar system analogs to guide us, these sub-Neptune objects remain shrouded in mystery. A key first step in determining the nature of sub-Neptune objects is estimating their masses, which provides a measure of their bulk density when combined with radius information from transit observations. Unfortunately, many sub-Neptune hosting systems are inaccessible to mass measurements via the radial velocity method for a variety of reasons (e.g. brightness of the host star in the visible, stellar activity, or long planetary orbital periods). However, many sub-Neptunes have been found in multi-planet systems with orbital configurations that allow for their masses to be estimated by measuring variations in their times of transit \citep{Schneider2003}. Systems hosting multiple sub-Neptune type planets also allow for a rich array of comparative studies that explore the formation and evolution of such planetary systems. 

The star within the K2-21 (EPIC 206011691) system is a relatively bright, nearby K7 dwarf \citep{Dressing2017} that hosts two transiting planets.  The formation and dynamical evolution of planetary systems orbiting K dwarfs is not well understood. To better understand the origin of compact systems of exoplanets, the first step in K2-21 requires robust constraints on the planetary masses, radii, and orbital eccentricities of both planets K2-21b and K2-21c. The initial photometric observations of these planets uncovered precise orbital periods of 9.32414 days and 15.50120 days, respectively, and a period ratio of 1.6624, which is relatively near the 5:3 mean motion resonance \citep{Petigura2015}.  This system's proximity to the mean motion resonance (MMR) is important because it affects the measured transit times to produce so-called transit timing variations (TTVs), where the gravitational perturbations between planets affect the transit times \citep{Agol2005,Holman2005}. The TTVs allow for the measurement of both the planet masses and their orbital eccentricities, which are better characterized when both planets are into or close to mean motion resonance. 
Here we present an analysis of the K2-21 system using observations from K2 and{\texttt{ Spitzer}}. The precision and temporal baseline of transit observations of both K2-21b and K2-21c from these facilities increase the fidelity of the mass and radius measurements of these ideal targets for comparative planetology studies. 

In the following sections we first describe the set-up of our observations and data reductions including observations from{\texttt{ Spitzer}} and K2 ($\S$\ref{sec_observations}), the set-up of our numerical model ($\S$\ref{sec_model}). In $\S$\ref{sec_results} we present the results of our investigation in terms of mass and orbital element determinations for both planets, and their possible compositions. Finally, in $\S$\ref{sec_conclusion} we present a summary of our results and discuss future works.

\section{Observations and Data Reduction} \label{sec_observations}

\begin{table}
  \caption{Transit times of planet K2-21b and K2-21c obtained by K2 and{\texttt{ Spitzer}} over a span of 4.9 years.}
  \label{tab_TT_obs}
\begin{tabular}{ccccc}
    \hline
Planet  & Transit   & Transit Time  &  $t_{\rm err}$ & Mission \\
        & Number    & (BJD - 2,456,900) & (Days)     &    \\
    \hline
b	&	 	 	1	&	80.09653	&	0.00492	 	 	&	K2	 	\\
b	&	 	 	2	&	89.41035	&	0.00692	 	 	&	K2	 	\\
b	&	 	 	3	&	98.75049	&	0.00494	 	 	&	K2	 	\\
b	&	 	 	4	&	108.0739 	&	0.00475	 	 	&	K2	 	\\
b	&	 	 	5	&	117.40152	&	0.00504	 	 	&	K2	 	\\
b	&	 	 	6	&	126.71487	&	0.00638	 	 	&	K2	 	\\
b	&	 	 	7	&	136.04625	&	0.00407 		&	K2	 	\\
b	&	 	 	8	&	145.37533	&	0.00544	 	 	&	K2	 	\\
b	&	 	 	49	&	527.74541	&	0.00212	 	 	&	{\texttt{ Spitzer}}	 	\\
b	&	 	 	72	&	742.27637	&	0.00268	 	 	&	{\texttt{ Spitzer}}	 	\\
b	&	 	 	89	&	900.80217 	&	0.00509	 	 	&	{\texttt{ Spitzer}}	 	\\
b	&	 	 	114	&	1133.91504 	&	0.00229	 	 	&	{\texttt{ Spitzer}}	 	\\
c	&	 	 	1	&	88.47122	&   0.00310	    	&	K2	 	\\
c	&	 	 	2	&	103.97430	&	0.00260	 	 	&	K2	 	\\
c	&	 	 	3	&	119.47245  	&	0.00258 	 	&	K2	 	\\
c	&	 	 	4	&	134.97949 	&	0.00239	 	 	&	K2	 	\\
c	&	 	 	29	&	522.45101	&	0.00161	 	 	&	{\texttt{ Spitzer}}	 	\\
c	&	 	 	43	&	739.42816	&	0.00280	 	 	&	{\texttt{ Spitzer}}	 	\\
c	&	 	 	54	&	909.92891	&	0.00227	 	 	&	{\texttt{ Spitzer}}	 	\\
c	&	 	 	78	&	1281.95023	&	0.00147	 	 	&	{\texttt{ Spitzer}}	 	\\
c	&	 	 	115	&	1855.39811	&	0.00178	 	 	&	{\texttt{ Spitzer}}	 	\\
 \hline
  \end{tabular}
\end{table}

\subsection{K2}\label{subsec_k2}

The Kepler Space Telescope observed K2-21 during Campaign 3 of the K2 mission, after the telescope lost its second of four reaction wheels. During K2 operations solar radiation pressure torqued the spacecraft about its barrel, which caused stars to drift by $\sim$1 pixel over 6-hour timescales. These drifts are corrected with periodic thruster firings \citep{Howell2014}. Due to the field rotation, raw aperture photometry from K2 contains systematic variability of $\sim$1$\%$ due to inter- and intra-pixel sensitivity variations and variable aperture losses. 

We correct for these systematics using the \texttt{everest2.0} code by \citet{Luger2018}. This code models stellar variability and instrumental systematics simultaneously with a Gaussian process plus a linear combination of the products of the fractional fluxes in each pixel up to third order. The coefficients in the linear model are determined through a regularized L2 optimization. When transits are present, the Gaussian process partially fits out the transit. Following the recommendation by \citet{Luger2018}, we mask out the transits and re-optimize \texttt{everest2.0}'s systematic and variability model. We subtract this model to produce a whitened light curve. The long-cadence integration comprising the whitened light curve had a dispersion of $\sigma$ = 1.48*MAD = 109ppm, which we adopt as the per-measurement uncertainty in fractional flux, where MAD denotes the mean absolute deviation. We use the \texttt{exoplanet} code \citep{Foreman-Mackey2013} to fit a two planet model \citep[i.e.,][]{Mandel2002} to the whitened photometry, using transit parameters from \citet{Petigura2015} as starting values. We then remove outliers that differ from the maximum {\it a posteriori} model by 5$\sigma$ or more.

\subsection{{Spitzer}}\label{subsec_spitzer}

We observe the K2-21 system with{\texttt{ Spitzer}} over a span of 3.5 years as part of four separate programs (11026, PI: Werner; 12035, 13045, 14244, PI: Stevenson).
Since the orbital solution was poorly constrained at the time of execution, program 11026 uses 14.6-hour observations to recover one transit of each planet.  Subsequent observations span 8.2 hours.  All programs use the 4.5 {\micron} channel with a 32$\times$32 subarray and 2-second frame times.  Each observation utilizes a $\sim$30-minute astronomical observation request (AOR) to settle the telescope pointing before reacquiring the target for the science AOR.

For all programs, we reduce the{\texttt{ Spitzer}} data using the Photometry for Orbits, Eclipses, and Transits (POET) pipeline \citep{Stevenson2011, Cubillos2013}.  Briefly, POET reads in the Basic Calibrated Data (BCD) files from by the{\texttt{ Spitzer}} Heritage Archive, identifies and masks bad pixels using a double-iteration, 4$\sigma$ outlier routine, computes centroids using a 2D Gaussian on each frame, and performs aperture photometry at various sizes.  For consistency, we use the same configuration for all observations (2.5 pixel aperture for target flux, 7-15 pixel annulus for background flux).
We fit each light curve individually using a transit model, a self-calibrating BiLinearly-Interpolated Subpixel
Sensitivity BLISS map for position-dependent systematics \citep{Stevenson2011}, and a linear trend in time.  We compute uncertainties using Differential-Evolution Markov Chain (DEMC) Monte Carlo techniques \citep{terBraak2006, terBraak2008}.

Spanning all four{\texttt{ Spitzer}} programs, we obtain four  transits the inner planet b and five transits of the outer planet c.  We definitively detect transits in nine of these datasets; the final observation of K2-21b (2019-10-23, AOR 69555456) exhibits a significant amount of red noise and the transit time is unconstrained.  Thus, we exclude this dataset from our remaining analyses.
The nine good{\texttt{ Spitzer}} transit times are listed in \Cref{tab_TT_obs}. 

At 4.5 {\micron}, we compute weighted mean planet-to-star radius ratios of $0.0307{\pm}0.0011$ and $0.0357{\pm}0.0009$ for planets b and c, respectively.  Using the GAIA-measured stellar radius of $0.5762^{+0.0332}_{-0.0397}$ $R_{\sun}$, the K2-21 planets have sizes of $1.93{\pm}0.07$ and $2.25{\pm}0.05$ $R_{\earth}$.

\subsection{{Least-Squares Fit}}\label{subsec_lsq}

We perform a least-squares minimization of the measured TTVs to fit a sinusoidal function with linear ephemeris of the form:  
\begin{equation} \label{eqn:TTV}
    \textrm{TTV} = A\sin{(2\pi(t-t_{\textrm{offset}})/p_{\textrm{super}})} + T_0 + np
\end{equation}
for each planet.  Here, $t$ is the measured transit time, $n$ is the known transit number, $T_0$ is the fitted mean ephemeris, $t_{\textrm{offset}}$ is the offset time, and $p$ is the fitted mean orbital period. Using both the Spitzer and K2 transit times, we find a best-fit super period of $p_{\textrm{super}}$ = 988.7 days and TTV semi-amplitudes $A$ of $39.7$ and $27.7$ minutes for planets b and c, respectively.  \Cref{tab:bestfitOC} contains all of our best-fit parameters from the fit.

\begin{table}
    \centering
    \begin{tabular}{ccc}
        \hline
        Parameter           & Value     & Unit      \\
        \hline
        Period of b         & 9.325898  & Days      \\
        $T_0$ of b          & 80.106039 & MJD       \\
        $A_b$               & 39.7      & Minutes   \\
        Offset of b         & 531.4     & Days      \\
        Period of c         & 15.499554 & Days      \\
        $T_0$ of c          & 88.464776 & MJD       \\
        $A_c$               & 27.7      & Minutes   \\
        Offset of c         & 24.3      & Days      \\
        $p_{\textrm{super}}$& 988.7     & Days      \\
        \hline
    \end{tabular}
    \caption{Best-fit parameters of planets K2-21b and K2-21c using both the Spitzer and K2 transit times (MJD = BJD - 2,456,900)}. 
    \label{tab:bestfitOC}
\end{table}

\begin{figure*}
\begin{subfigure}{.5\textwidth}
  \centering
  \includegraphics[width=1.\linewidth]{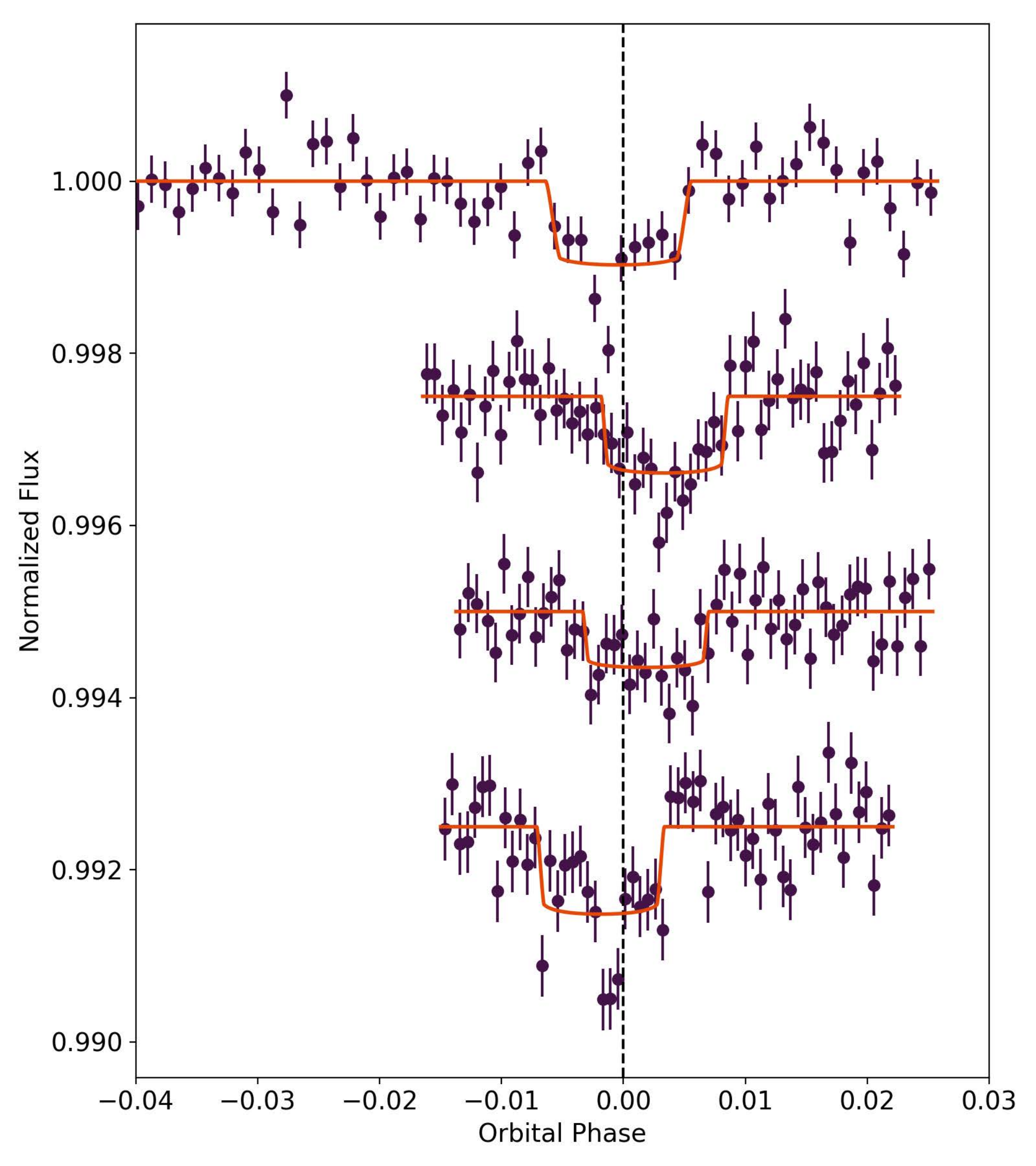}
  \label{fig:LCb}
\end{subfigure}%
\begin{subfigure}{.5\textwidth}
  \centering
  \includegraphics[width=1.\linewidth]{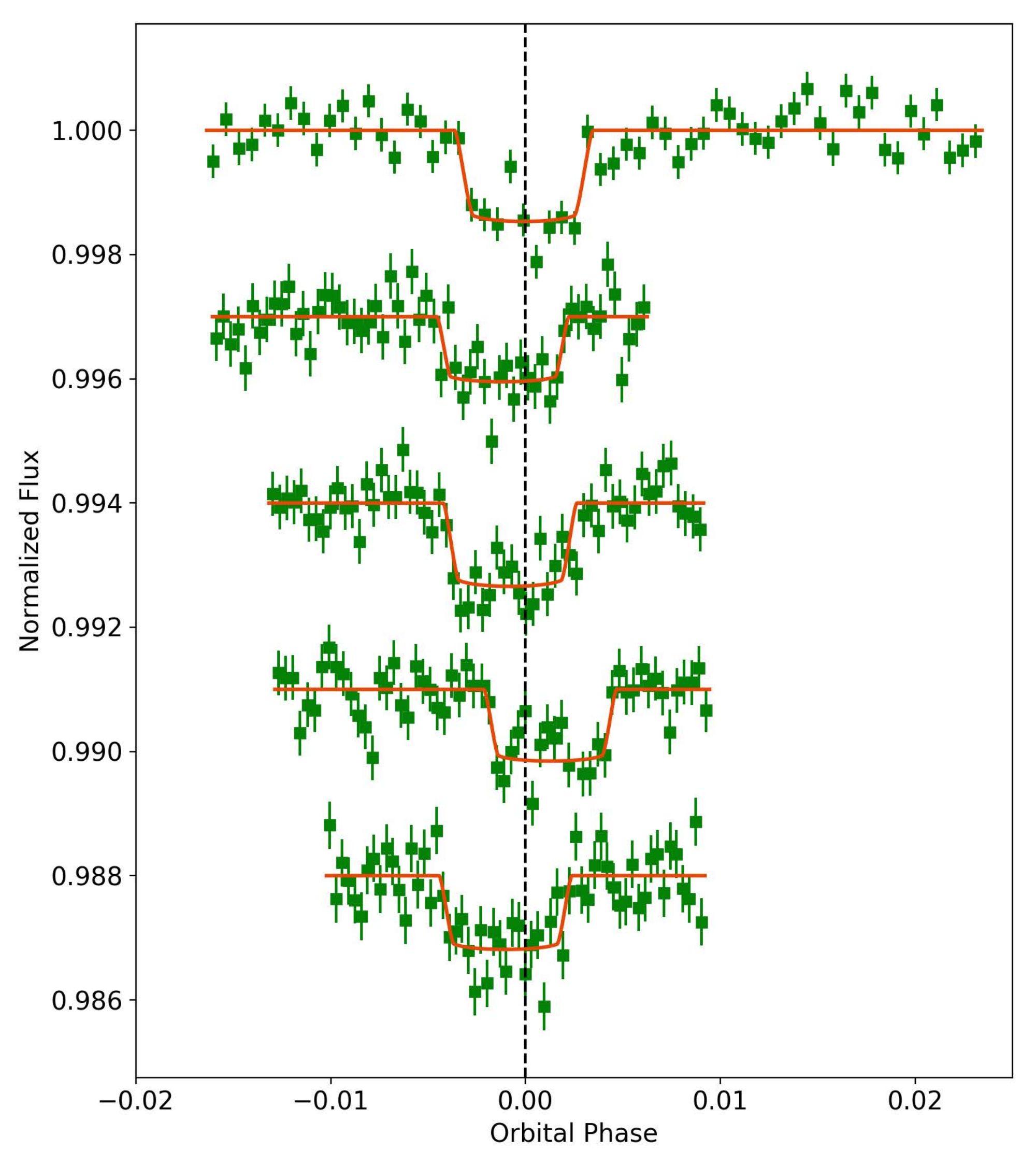}
  \label{fig:LCc}
\end{subfigure}
 \caption{{\texttt{ Spitzer}} transit observations of K2-21b (left) and K2-21c (right) with best fit models. The transit timing variations range from -40 to +40 minutes for K2-21b and from -28 to +28 minutes for K2-21c.}
\label{fig_LCb_LCc}
\end{figure*}

\Cref{fig_LCb_LCc} shows the{\texttt{ Spitzer}} transit lights curves for planets b (left) and c (right), respectively.  We use the mean orbital periods (9.325898 and 15.499554 days) and ephemerides (80.106039 MJD and 88.464776 MJD) to compute each static orbital phase along the abscissa ($x$ axis).  The TTVs are clearly seen by eye.
The mean orbital period ratio between the two planets is 1.661991, which is slightly less than the 5:3 orbital resonance. 
The top panel of \Cref{fig_OC_chop} represents the observed minus calculated diagram for K2-21b and K2-21c fitted using a simple Lavenberg-Marquardt fit.

\section{Transit Timing Methods and Analysis} \label{sec_model}
To estimate the planetary parameters in K2-21, we implement a forward model using the Python wrapper version of \texttt{TTVFast} \citep{Deck2014} that uses the stellar mass $M_\star$ and seven parameters for each planet (mass $M$, period $P$, $e\cos\omega$, $e\sin\omega$, inclination $i$, nodal longitude $\Omega$, \& time of first transit $T_o$) as input to evaluate the resultant transit times, where $e$ and $\omega$ are the orbital eccentricity and longitude of the pericenter, respectively. The \texttt{TTVFast} module propagates the model relative to a reference epoch $T_{\rm ref}$ over the range of transits given in \Cref{tab_TT_obs} using a timestep of 0.045 days (about $0.5\%$ of the orbital period of the inner planet) in a Jacobian coordinate system \footnote{A listing of the requirements are described in the {\texttt{Fortran}} version of \texttt{TTVFast}}.  \texttt{TTVFast} returns a matrix of results denoting the planet index, transit number and transit time in days.  It also provides the position and velocity of the planets on the sky so that radial velocity measurements can be included in the model, but that portion of the module is not used in our work. 

To efficiently survey the parameter space, we implement the commonly used \texttt{emcee} package \citep{Foreman-Mackey2013}. The sampling routines within \texttt{emcee} are frequently used for many astrophysical problems including the estimation of planetary parameters from a list of transit times. We modify the \texttt{emcee} sampler to use differential evolution proposals \citep{Nelson2014} along with a snooker updater \citep{terBraak2008}, which improved its performance.  The \texttt{emcee} sampler uses a set of assumed initial parameters (\Cref{tab_results}) along with their normally distributed uncertainties $\sigma$ to generate a large set of walkers that explore the parameter space.  These walkers are modified through our $\log{\chi^2}$ likelihood function, when the transit times determined from the model are compared with the observations (\Cref{tab_TT_obs}), or  
\begin{equation}
    \mathbf{{\log{\chi^2}}} = -\frac{1}{2} \sum_{n=1}^{N} \left(   \frac{O^n - C^n}{\sigma}\right)^2 + \log \left(2\pi\sigma^2\right), 
\end{equation}
where $N$ is the total number of transit times, and $O^n$ and $C^n$ are the observed and calculated n-th transit time, respectively. 

As a result, the \texttt{emcee} package traverses sets of parameters along the path of largest probability. To better constrain the fitting procedure, we use a prior probability function based on prior results from \citet{Petigura2015} to limit the exploration of parameter space by \texttt{emcee}, where values outside the allowed range receive a penalty. For example, we penalize the trial parameters that would be nonphysical (e.g., using the planet packing conditions from \citet{Gladman1993}). To narrow the phase parameter to get the best solutions, we add an additional prior that checks whether the trial parameters produce an impact parameter \citep{Winn2010} that is consistent (within 1$\sigma$) with the observations. The impact parameter $b$ is the sky-projected distance at conjunction. In this case, we limited the code to allow solutions where $0.25<b<0.85$, using $b_b = 0.7^{+0.15}_{-0.45}$ and $b_c = 0.66^{+0.19}_{-0.40}$ for the inner and outer planet, respectively \citep{Petigura2015}. Our code is available, upon request from the first author, in the hosting platform \texttt{GitHub} via this link: https://github.com/MaryameElMoutamid/Exoplanets/TTV.

\section{Results} \label{sec_results}

\subsection{Parameters and possible Compositions of the two K2-21 Planets}\label{subsec_results_composition}

\begin{figure*}
\includegraphics[width=0.993\linewidth]{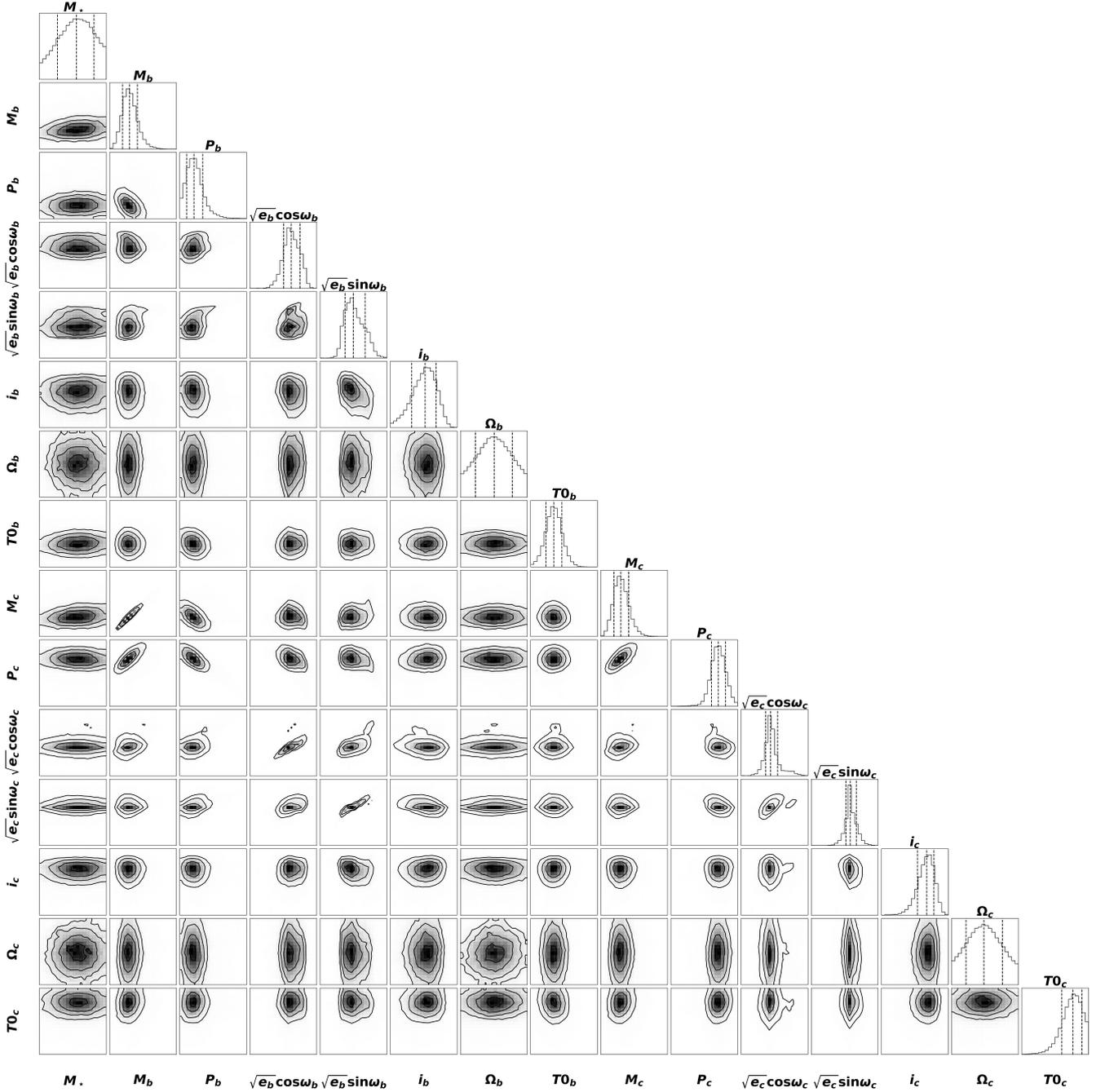}
\caption{Posterior distribution of all 15 free parameters (the star's mass and 6 parameters for each planet) using our model described in \Cref{sec_model} providing the most likely solution for the masses and orbital elements of both planets in the K2-21 system.}
\label{fig:corner_complete}
\end{figure*}

\begin{figure}
\includegraphics[angle=-90,width=9cm]{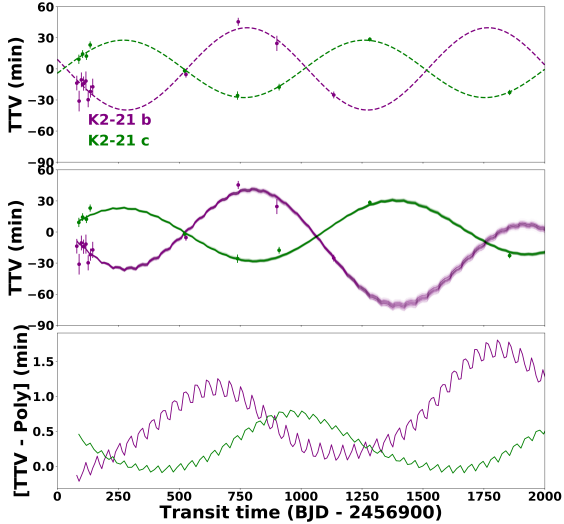}
\caption{Top: Observed minus calculated diagram using \Cref{eqn:TTV} for K2-21b and K2-21c fitted with the best fit using a simple Lavenberg-Marquardt fit.
 Middle: Transit timing variation (TTV) measurements and uncertainties for K2-21b (purple) and K2-21c (green) using observations from K2 and Spitzer. The solid curves illustrate the TTV using posterior samples from our dynamical model with \texttt{TTVFast}. This model diverges from the one represented in the top panel after $\sim$700 days because \texttt{TTVFast} accounts for more physical parameters, such as the planetary eccentricity. Bottom: Short-timescale chopping variations of the best fit model. 
 \label{fig_OC_chop}}
\end{figure}

\begin{figure}
 \includegraphics[width=1.1\columnwidth]{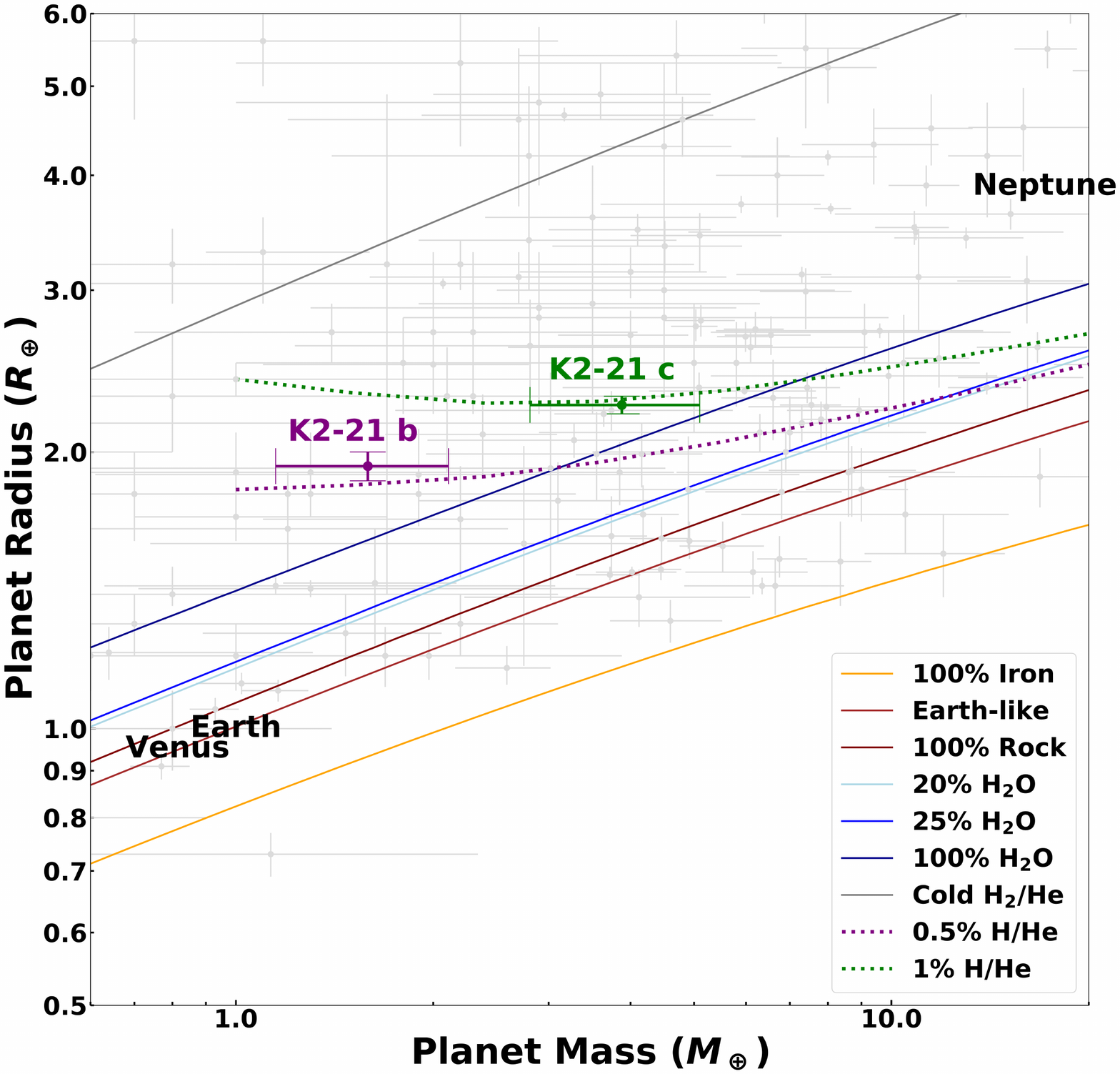}
 \caption{Mass-radius logarithmic diagram using our mass constrains and radius uncertainties from {\texttt{Spitzer}} analysis. Three planets of the Solar System have been added for illustration in black. The solid curves indicate expected planet mass-radius curves for different compositions, by interpolating masses and radius from the third table \citet{Zeng2016}. The dotted curves illustrate the mass-radius relationship using the first table in \citet{Lopez2014}. Gray measurements on the background represent results on mass-radius of exoplanets reported by \citet{Orosz2019}.}
 \label{fig_density}
\end{figure}

For both inner and outer planets, we report masses of $1.59^{+0.52}_{-0.44}\ M_\oplus$ and $ 3.88^{+1.22}_{-1.07}\ M_\oplus$ and densities of $ 0.22^{+0.05}_{-0.04}\ \rho_\oplus$ and $0.34^{+0.08}_{-0.06}\ \rho_\oplus$, respectively, where $M_\oplus$ and $\rho_\oplus$ are the mass and density of Earth, respectively. \Cref{fig:corner_complete} and \Cref{tab_results} summarize our results on the masses and orbital elements of these 2 planets.  
These values are verified from the short-timescale chopping variations (\Cref{fig_OC_chop}). These variations are caused by coherent small perturbations at conjunctions that could change the planet orbits \citep{Lithwick2012,Nesvorny2014,Deck2015,Hadden2016,Agol2021} of the best fit model.  
By combining these new constraints on the masses of the two K2-21 planets with the measured radii by \citet{Petigura2015}, we provide constraints on their possible bulk densities. \Cref{fig_density} shows the two K2-21 planets and the theoretical mass‚ radius functions for different bulk planetary densities from \citet{Zeng2016} and \citet{Lopez2014}.
Both planets would have properties consistent of a hydrogen and helium dominated atmospheres.

However, because of the significant mass uncertainties, the two planets could also have a substantial fraction of their masses in large volatile envelopes formed from water or other volatile ices.
We report that planet b has a lower density and thus, puffier than planet c. This suggests that the K2-21 planetary system may not have formed \textit{in situ} like the Galilean satellites because planet b would have suffered a stronger atmospheric erosion through hydrodynamic escape given its proximity to the host star.  More observations are needed to better constrain formation scenarios \citep{Sotzen2021,Benneke2013}.
\begin{table*}
  \centering
  \caption{Modeling planetary parameters for the K2-21 system. All $T_0$ values are provided on BJD - 2,456,900. The mass of the K2-21 star is $0.676023$ $M_{\sun}$, and the upper and lower values of the stellar mass that we use in our code are $0.743625$ $M_{\sun}$ and $0.608421$ $M_{\sun}$, respectively}. 
  \label{tab_results}
  \begin{tabular}{|l|ll|l|ll|l|}
    \hline
    \xrowht{10.5pt}
    {Planets} & \multicolumn{3}{c|}{K2-21b} & \multicolumn{3}{c|}{K2-21c}   \\
    \hline
    \xrowht{10.5pt}
    \multirow{2}{*}{Parameters} &  \multicolumn{2}{c|}{Uniform prior bound} & \multirow{2}{*}{Outputs} & \multicolumn{2}{c|}{Uniform prior bound} & \multirow{2}{*}{Outputs}    \\ \cline{2-3} \cline{5-6}
    \xrowht{10.5pt}
        &    lower &  upper  &  &  lower &  upper  & \\
    \hline
    Mass (M$_\oplus$)     & $0.1$ & $30.$   & $1.59^{+0.52}_{-0.44}$ &  $0.1$ & $30.$   & $3.88^{+1.22}_{-1.07}$ \\
    \xrowht{12.pt}
    P (days)              & $ 9.32351$  & $ 9.32473$    &  $9.3238^{+0.0002}_{-0.0001}$ &  $15.50021 $ & $ 15.50213$    & $15.5017^{+0.0002}_{-0.0002}$ \\
    \xrowht{12.pt}
    $\sqrt{e}\cos\varpi$  & $-0.7 $ & $ 0.7$    &  $-0.03^{+0.14}_{-0.12}$ &  $-0.7$ & $0.7$   &  $-0.18^{+0.09}_{-0.06}$\\
    \xrowht{12.pt}
    $\sqrt{e}\sin\varpi$  & $-0.7 $ & $ 0.7$    &  $0.04^{+0.21}_{-0.15}$ & $-0.7 $ & $ 0.7$    &  $0.15^{+0.09}_{-0.07}$ \\
    \xrowht{12.pt}
    $i_{\text{Sky}}$ ($^{\circ}$)  & $ 87.$ & $ 89.99 $    & $ 88.54^{+0.49}_{-0.59}$ & $ 87.$ & $ 89.99 $    &  $ 89.02^{+0.33}_{-0.41} $ \\
    \xrowht{12.pt}
    $\Omega$ ($^{\circ}$)  & $-0.5 $ & $ 0.5$    & $0.0048^{+0.2746}_{-0.2774}$  & $-0.5 $ & $ 0.5$    & $ -0.0141^{+0.2772}_{-0.2687}$ \\
    \xrowht{12.pt}
    $T_0$ (days)         & $80.0942$ &  $80.11734$   & $80.0977^{+0.0012}_{-0.0012}$ &  $88.47122$ & $88.47466 $    & $88.4739^{+0.0005}_{-0.0006}$ \\
        \hline 
  \end{tabular}
 \end{table*}

\subsection{On the formation, stability and resonant configuration of the K2-21 system}\label{subsec_results_stability}
The two planets that we model in the K2-21 system are similar in density to Jupiter or Neptune, where there are competing theories as to the origins of the low mass planets with large radii \citep{Jontof-Hutter2019}.  For such planets with TTV derived masses, the typical origin would be through Type I migration where the planets begin just beyond the snow line, develop into a MMR, and migrate inward while maintaining the resonant configuration.  However, \cite{Lee2016} suggested that super-Puffs (4--10 R$_\oplus$, 2--6 M$_\oplus$) form within a gas-poor region of the disk ($\sim$1 au) that coincides with a large shift in the conventional snow line during the pre-main-sequence \citep{Kennedy2008}.

The long-term stability of two planet systems like K2-21 is not well defined.  \cite{Petigura2015} only estimate the lower limit for the stellar age $\gtrsim$1 Gyr, where the system could be in a meta-stable state and the age of the system cannot be used as a constraint.  Our TTV modeling penalizes trials where the planetary spacing is less than 2$\sqrt{3}$R$_{\rm H}$ \citep{Gladman1993} and thus our results could represent stable two-planet systems in the short-term ($\lesssim 1000$ yr).  Previous studies \citep{Chambers1996,Smith2009,Pu2015,Goldberg2022} showed that planetary spacing parameters for systems of multiple planets lie between $4-10$ depending planet multiplicity, where the most compact systems (e.g., TRAPPIST-1; \citet{Quarles2017}) $\sim$10 mutual Hill radii of space between adjacent pairs of planets.  However, the planetary spacing derived from our posterior is 21.5 $\pm$ 1.6 mutual Hill radii, which permits planetary stability for low eccentricity orbits. Instabilities due to the overlap of first order mean motion resonances \citep{Deck2013} are not applicable because the K2-21 system lies near a second order mean motion resonance (5:3) and the resonance overlap typically occurs for period ratios inside the 4:3 \citep{Lithwick2012}. 

To test the potential for longer term stability, we use the posterior parameter distribution from our TTV modeling and perform numerical simulations for $10^6$ orbits of the inner planet ($\sim25,000$ yr) using the n-body simulation package \texttt{REBOUND} \citep{Rein2012,Rein2015}.  To balance the computational expense with numerical accuracy, we use the \texttt{whfast} integration scheme with an 11th order symplectic corrector, where the integration timestep is 5\% of planet b's orbital period ($\sim$0.46 days).  From the simulation of 3000 samples, we find that all of our TTV solutions are stable for $10^6$ orbits of the inner planet, which is $\sim$5 times longer than the secular forcing timescale ($\sim$5000 yr).  Instabilities could occur on much longer timescales, but other physical effects (e.g., tides and/or General Relativity) need to be included before concluding on the stability of the K2-21 system. 

Many of the TTV systems identified in the Kepler data have been analyzed to determine the period ratios of adjacent planets and their proximity to mean motion resonances \citep{Fabrycky2014}.  We calculate the proximity to the 5:3 resonance using our posterior through the following relation \citep{Hadden2016},
\begin{equation}
    \Delta = \frac{P_{\rm c}}{P_{\rm b}}\frac{K-2}{K}-1,
\end{equation}
where $K=5$ and find $\Delta\approx -0.00242$.  Finding a system this close to a mean motion resonance is definitely intriguing, but there is a secondary condition for resonances which includes the libration of its resonant angle $\phi$.  The resonant angle depends on a linear combination of the mean longitudes ($\lambda_b,\:\lambda_c$), the arguments of pericenter ($\varpi_b,\:\varpi_c$) and the longitudes of the ascending node ($\Omega_b$,$\Omega_c$) of the two planets.  

We examine the following conditions for 6 possible 5:3 eccentric resonances, using the values at a given epoch and using 5000 day n-body simulations to test whether the resonant angle librates around $\phi\approx0^\circ$ or $\phi\approx180^\circ$. 
\begin{equation}
\begin{cases} 
\phi_1 = 5\lambda_c - 3\lambda_b - 2\varpi_b, \\ 
\phi_2 = 5\lambda_c - 3\lambda_b - 2\varpi_c,  \\
\phi_3 = 5\lambda_c - 3\lambda_b - \varpi_b - \varpi_c, \\
\phi_4 = 5\lambda_c - 3\lambda_b - 2\Omega_b, \\ 
\phi_5 = 5\lambda_c - 3\lambda_b - 2\Omega_c, \\ 
\phi_6 = 5\lambda_c - 3\lambda_b - \Omega_b - \Omega_c
\end{cases}
\label{eq_angles}
\end{equation}

All of the temporal variations of the 6 resonant angles from \Cref{eq_angles} at $T_{\rm ref} = 2,456,975$ days showed that K2-21b and K2-21c are not in resonance. 

\section{Conclusions} \label{sec_conclusion}

\begin{figure*}
\begin{subfigure}{.5\textwidth}
  \centering
  \includegraphics[width=1.\linewidth]{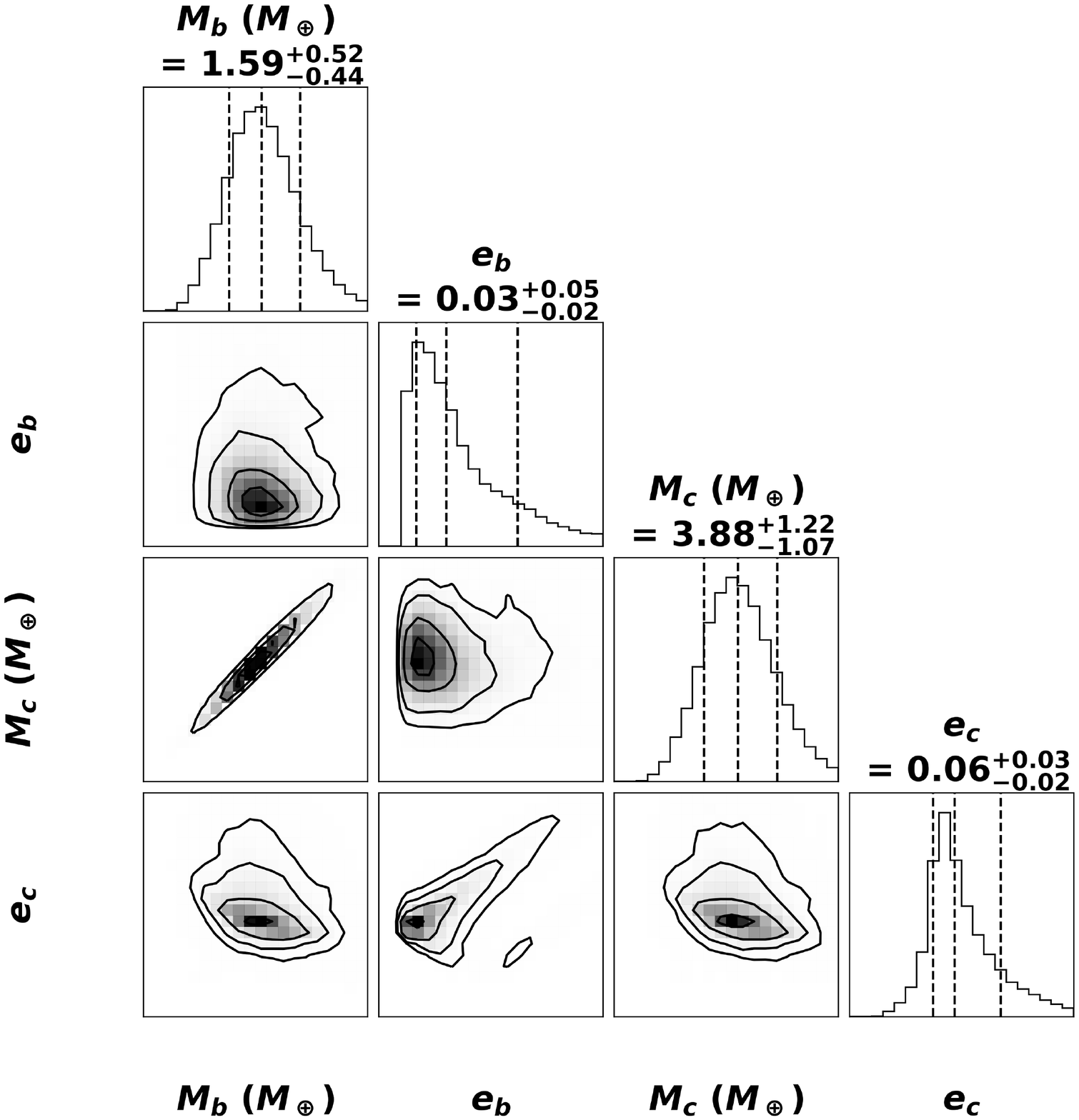}
  \label{fig:fig_corner_me_1}
\end{subfigure}%
\begin{subfigure}{.5\textwidth}
  \centering
  \includegraphics[width=1.\linewidth]{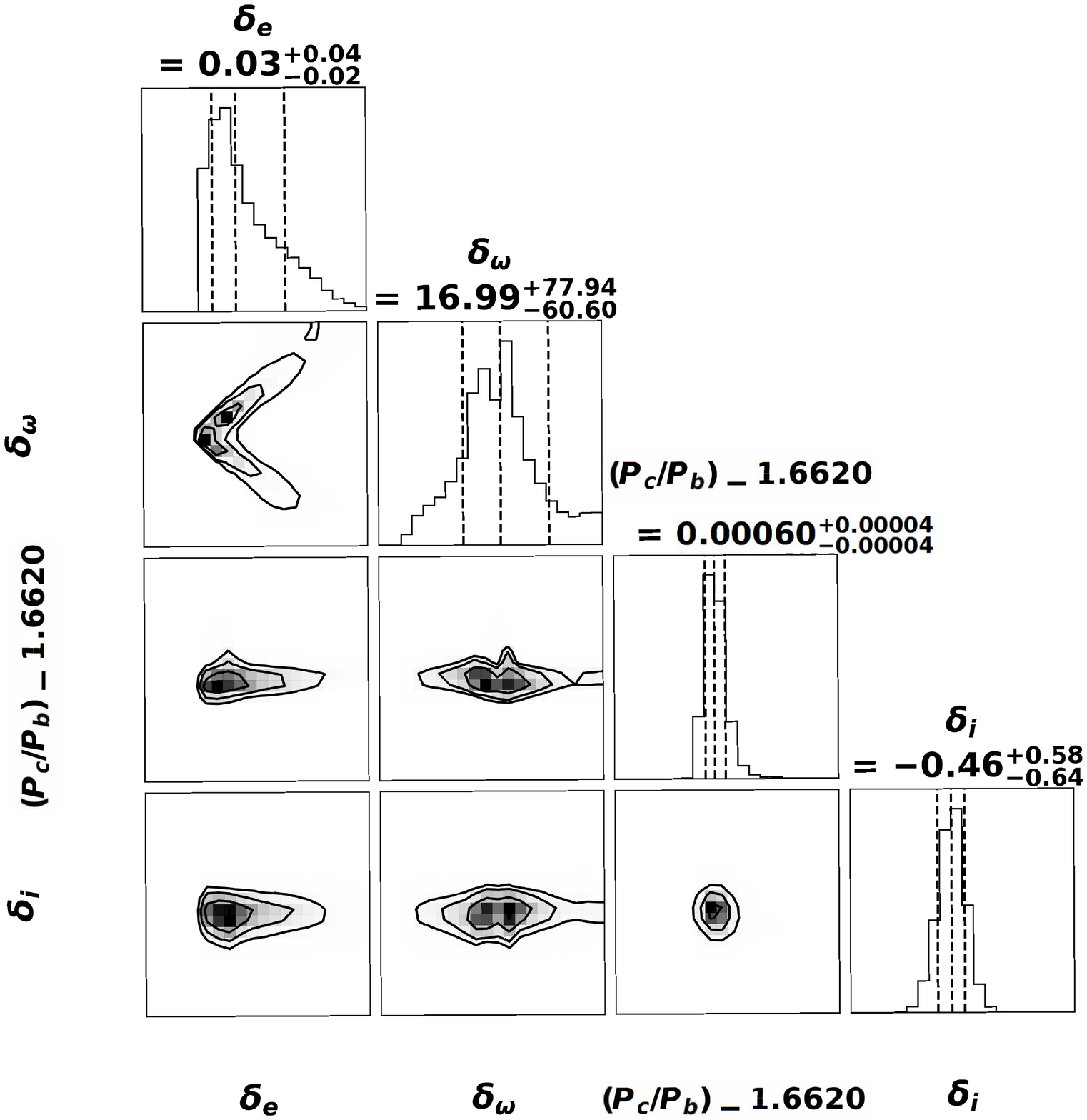}
  \label{fig:fig_corner_me_4}
\end{subfigure}
 \caption{Left panel: Posterior samples of the planet masses ($M_b$ and $M_c$) and eccentricities ($e_b$ and $e_c$) from our numerical model. Right panel: Posterior samples of the difference in eccentricity vector components $\delta e$, the difference of longitudes of pericenters, the difference of planet period ratio from 1.662 \citep{Bailey2022} and the difference of inclinations for both planets.}
\label{fig:fig_corner_me}
\end{figure*}

From the transit timing variations (TTVs) of K2-21, we derive the masses and orbital eccentricities of both planets K2-21b and K2-21c. We find that the inner and outer planets in the K2-21 system have interiors consistent with the presence of a hydrogen and helium dominated atmosphere from our estimate of their masses, which are $1.59^{+0.52}_{-0.44}\ M_\oplus$ and $ 3.88^{+1.22}_{-1.07}\ M_\oplus$ and densities of $ 0.22^{+0.05}_{-0.04}\ \rho_\oplus$ and $0.34^{+0.08}_{-0.06}\ \rho_\oplus$, respectively. 
Our results show that the inner planet is less dense than the outer planet; one more counter-intuitive exoplanetary system such as Kepler-105 \citep{Wang2014,Rowe2014}, LTT 1445 \citep{Winters2019,Winters2022}, TOI-175 \citep{Kostov2019,Demangeon2021} and Kepler-279 \citep{Xie2014,Rowe2014} systems. 
Both planets are beyond the transition radius of $1.6\ R_\oplus$ and are expected to be low density from mass-radius relations \citep{Rogers2015,Chen2017}. We find that these two planets are near (but not in) a second-order mean motion resonance (5:3) on eccentric orbits. Indeed, the observed transit periods are not oscillating around the exact ratio 1.6667 \citep{Bailey2022}, suggesting that these two planets may not be in resonances as previously thought. 
Although, the measured period ratio is 1.6626 $\pm{0.0002}$, which is consistent (at $2\sigma$) with the value reported by \citet{Petigura2015} of 1.6625. We have added posteriors and joint posteriors for the difference in eccentricity
$\delta e$, longitudes of pericenter $\delta \omega$, and inclinations $\delta i$, as a companion corner plot (\Cref{fig:fig_corner_me}). 

The James Webb Space Telescope (JWST) would provide additional observations of the K2-21 system, which would allow for further optimization of the main parameters of this system, by adding transit and eclipse times for both planets. With these new observations, we expect improvements in the accuracy of observational parameters, such as the transit depth, transit duration, and impact parameter for each planet. Such additional measurements would also affect the prior mass measurements via TTVs. In fact, the longer baseline (i.e., gap between measurements) could narrow the parameter space of both mass and eccentricity to improve the fit. JWST could also refine the orbital resonant configuration of the K2-21 system.

\section*{Acknowledgements}
We acknowledge resonant TTV modeling kindly provided by Dr. Sean Mills was used in the Spitzer proposals. This work is based [in part] on observations made with the Spitzer Space Telescope, which was operated by the Jet Propulsion Laboratory, California Institute of Technology under a contract with NASA. This research was supported in part through research cyberinfrastructure resources and services provided by the Partnership for an Advanced Computing Environment (PACE) at the Georgia Institute of Technology.  The authors would like to thank Dr. Laura Kreidberg for helpful discussions. Dr. Diana Dragomir acknowledges support from the TESS Guest Investigator Program grant \verb|80NSSC21K0108| and NASA Exoplanet Research Program grant \verb|18-2XRP18_2-0136|. The authors thank the anonymous reviewer for their useful remarks and suggestions.

\section*{Data Availability}
Data are available, upon request from the first author, in a repository and can be accessed via the DOI link https://github.com/MaryameElMoutamid/Exoplanets/TTV.

\clearpage
\newpage
\bibliography{ms}



\clearpage
\newpage

\onecolumn 
APPENDIX A: Code validation with synthetic data \\

\begin{figure}
\includegraphics[width=0.993\linewidth]{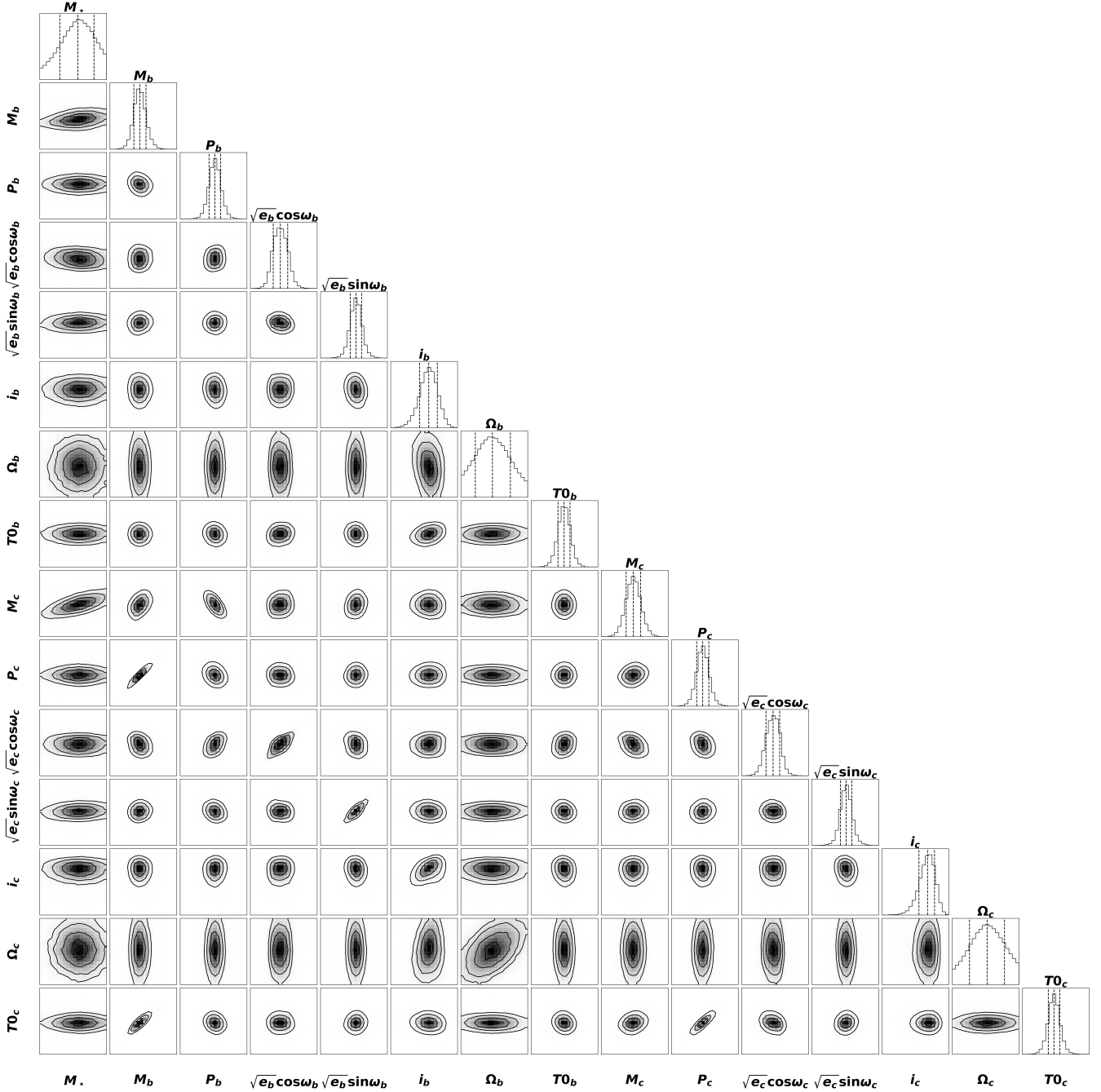}
\caption{Posterior distribution of all 15 free parameters (including the star's mass) using our model described in \Cref{sec_model} providing the most likely solution for the masses and orbital elements of synthetic data of a system with two planets with masses and orbital parameters listed in \Cref{tab_syn}.}
\label{fig:corner_syn}
\end{figure}

To evaluate the robustness of our code, we test it by using synthetic data produced from a set of known parameters (e.g., star mass, planetary masses and orbital parameters), and check if our code recovers these inputs. The initial parameters that we chose to reproduce and the output results are listed in the top of \Cref{fig:corner_syn}.  
Our synthetic transit times values are generated based on mean value of each synthetic planet parameters and the star mass, and from the posteriors of our model. We then add white noise to each transit time (up to 7 min). Our posterior results for masses and the orbital parameters are consistent with input in terms of their mean values. This test validates both our transit timing method and our TTV analysis, and allows us to use our code for real data in order to determine K2-21 planet parameters with confidence.

\begin{table}[h]
  \centering
  \caption{Modeling planetary parameters for a system with synthetic data close to K2-21 parameters. All $T_0$ values are provided on BJD - 2,456,900.}
  \label{tab_syn}
  \begin{tabular}{|l|l|lll|l|l|lll|l|}  
    \hline
    \xrowht{10.5pt}
    {Planets} & \multicolumn{5}{c|}{Planet b} & \multicolumn{5}{c|}{Planet c}   \\
    \hline
    \xrowht{10.5pt}
    \multirow{2}{*}{Parameters} & \multirow{1}{*}{True} & \multicolumn{3}{c|}{Uniform prior bound} & \multirow{2}{*}{Outputs} & \multirow{1}{*}{True} & \multicolumn{3}{c|}{Uniform prior bound} & \multirow{2}{*}{Outputs}    \\ \cline{3-5} \cline{8-10}
    \xrowht{10.5pt}
        & values & lower & guessed & upper  &  & values & lower & guessed & upper  & \\
    \hline
    Mass (M$_\oplus$) & 2.00  & $0.10$ & $4.00$ & $5.00$  & $ 1.98^{+0.38}_{-0.36} $ & 5.00 & 0.10 & 7.00 & 10.00   & $ 4.96^{+0.38}_{-0.38} $ \\   
    \xrowht{12.pt}
    P (days)  & 9.00314  & 9.00150 & 9.00314 & 9.00478 & $ 9.00315^{+0.00002}_{-0.00002} $ & 15.00 & 14.9994 & 15.00 & 15.0006  &  $ 15.0000^{+0.0001}_{-0.0001} $\\  
    \xrowht{12.pt}
    $\sqrt{e}\cos\varpi$ & 0.01  & -0.30 & 0.01 & 0.30 & $-0.005^{+0.051}_{-0.049} $ & 0.01 &  -0.30 & 0.01 & 0.30  & $0.008^{+0.041}_{-0.044} $\\  
    \xrowht{12.pt}
    $\sqrt{e}\sin\varpi$ & 0.01  & -0.30 & 0.01 & 0.30 & $0.005^{+0.043}_{-0.040} $ & -0.01 &  -0.30 & 0.01 & 0.30  & $0.002^{+0.041}_{-0.040} $ \\  
    \xrowht{12.pt}
    $i_{\text{Sky}}$ ($^{\circ}$)  & 88.30  & 87.00 & 88.30 & 89.99 & $ 88.66^{+0.37}_{-0.40} $ & 89.08 & 87.00 & 89.08 & 89.99 &  $ 89.21^{+0.25}_{-0.26} $\\  
    \xrowht{12.pt}
    $\Omega$ ($^{\circ}$) & 0.00  & -0.50 & 0.00 & 0.50 & $-0.03^{+0.27}_{-0.26} $ &  0.00 & -0.50 & 0.00 & 0.50 & $0.02^{+0.26}_{-0.26} $ \\  
    \xrowht{12.pt}
    $T_0$ (days) & 80.00  & 79.9983 & 80. & 80.0017  & $ 80.00001^{+0.00004}_{-0.00004} $ & 88.00 & 87.9983 & 88. & 88.0017 & $ 88.00001^{+0.00010}_{-0.00010} $  \\  
    \hline
  \end{tabular}
 \end{table}

\bsp	
\label{lastpage}

\end{document}